\documentstyle[preprint,pra,eqsecnum,aps]{revtex}
\begin{document} \draft

\title{Jones-matrix Formalism as a Representation of the Lorentz Group}

\author{D. Han\footnote{electronic mail: han@trmm.gsfc.nasa.gov}}
\address{National Aeronautics and Space Administration, Goddard Space
Flight Center, Code 910.1, Greenbelt, Maryland 20771}

\author{Y. S. Kim\footnote{electronic mail: kim@umdhep.umd.edu}}
\address{Department of Physics, University of Maryland, College Park,
Maryland 20742}

\author{Marilyn E. Noz \footnote{electronic mail: noz@nucmed.med.nyu.edu}}
\address{Department of Radiology, New York University, New York, New York
10016}

\maketitle

\begin{abstract}
It is shown that the two-by-two Jones-matrix formalism for polarization
optics is a six-parameter two-by-two representation of the Lorentz group.
The attenuation and phase-shift filters are represented respectively by
the three-parameter rotation subgroup and the three-parameter Lorentz
group for two spatial and one time dimensions.  It is noted that the
Lorentz group has another three-parameter subgroup which is like the
two-dimensional Euclidean group.  Possible optical filters having this
Euclidean symmetry are discussed in detail.  It is shown also that the
Jones-matrix formalism can be extended to some of the non-orthogonal
polarization coordinate systems within the framework of the Lorentz-group
representation.

\end{abstract}

\narrowtext

\section{Introduction}

The Lorentz group was introduced to physics as the language of
space-time symmetries of relativistic particles~\cite{wig39,barg47}.
However, this group serves useful purposes in many other branches of
physics, including optical sciences.  In recent years, the Lorentz
group served as the underlying language for squeezed states of light.
It was Dirac who first observed that the Lorentz boost is a squeeze
transformation~\cite{dir49}.  In 1963~\cite{dir63}, Dirac constructed
representations of the Lorentz group using coupled harmonic oscillators.
Indeed, Dirac's oscillator representation forms the theoretical
foundations of squeezed states of light~\cite{yuen76,cav85,knp91}.
This aspect of the Lorentz group is by now well known in the optics
community, and the Lorentz group is  one of the theoretical tools in
optics.

The squeezed state is not the only branch of optics requiring the
Lorentz group.  In 1981, Bacry and Cadihilac considered application of
the Lorentz group for Fourier optics~\cite{bacry81}.  In 1983,
Sudarshan {\it et al}. constructed a representation of the Lorentz
group for para-axial wave optics~\cite{sudar83}.  Wavelets are known to
be representations of the Lorentz group~\cite{hkn95}.  It has been
noted recently that the Lorentz group serves as the underlying language
for reflections and refractions~\cite{pellat91}.  In addition, in their
recent paper~\cite{gutier96}, Gutierrez {\it et al}. pointed out the
relevance of the Lorentz geometry to three-dimensional concentrators.

More recently, the relevance of Lorentz transformations to polarization
optics has been discussed by several
authors~\cite{pellat91,cloude86,chiao88,kitano89,perina93}.  Indeed, in
their recent paper~\cite{perina93}, Opartrny and Perina have shown that
rotation and boost matrices from the two-by-two representation of the
Lorentz group describe energy-conserving and non-conserving optical
filters respectively.  In our recent paper, we pointed out that
polarization optics can be formulated in terms of the six-parameter
Lorentz group~\cite{hkn96}.

The purpose of this paper is very simple.  The standard language for
optical polarizations is the Jones-matrix formalism, and it has a long
history~\cite{jones41,swind75,predo93}.  In this paper, we show that this
formalism is a representation of the six-parameter Lorentz group.
The Lorentz group also has a long history.

Among the many different representations of this group, as has been
shown in one of our earlier papers~\cite{hkn96}, the bilinear conformal
representation~\cite{barg47} is most convenient for polarization optics.
Unfortunately, however, we were not able to connect there the
representation with the Jones-matrix formalism, and we shall complete
the task in the present paper.

The six-parameter Lorentz group has three major subgroups.  They are
the three-parameter rotation group, the three-parameter $O(2,1)$ group,
and the three-parameter $E(2)$-like group.  The mathematics of the
rotation group is well known.  It is shown in Ref.~\cite{hkn96} that
the rotation group is the underlying language for optical filters
causing phase shifts
between the two transverse components.  The $O(2,1)$-like groups have
been extensively discussed recently in connection with squeezed states
of light, and it is a very familiar language in optics.  It is known
that this $O(2,1)$-like group is the proper language for attenuation
filters~\cite{kitano89,hkn96}.

The $E(2)$-like group is useful in understanding the internal space-time
symmetries of massless particles~\cite{hks82}, but is relatively new in
optics~\cite{ky92}.  We discuss in this paper possible optical filters
possessing this $E(2)$-like symmetry.  Since this requires a detailed
knowledge of the six-parameter Lorentz group, and since this group is
relatively new in optics, we shall include in this paper a systematic
introduction to the Lorentz group applicable to three spatial and one
time-like dimensions.

The rotation and squeeze transformations discussed in this paper are
directly applicable to coordinate transformations.  Thus, they are
applicable to the case of where the polarization occurs along a
skew or squeezed coordinate system.  It is known that one of the
$E(2)$-like transformation leads to a ``shear'' transformation.  The
formalism is therefore is applicable also to the case where the
polarization coordinate is sheared.  The polarization plane is not
always perpendicular to the propagation direction of the light wave.
The formalism can be extended to accommodate this case.

In Sec. \ref{formalism}, we prove that the Jones-matrix formalism is
indeed a representation of the Lorentz group.  In Sec. \ref{combine},
the mathematics of Sec. \ref{formalism} is translated into
transformation matrices
corresponding to optical filters.  It is shown that the combined effect
of attenuation and phase-shift filters leads to a six-parameter
two-by-two matrix.  It is noted in Sec. \ref{conform} that the
bilinear conformal representation of the Lorentz group is the natural
scientific language for polarization optics.  It shown that the
attenuation and phase-shift filters have their own sub-representations.
It is shown that one of those sub-representations leads to a new
kind of optical filter having the $E(2)$-like symmetry.
In Sec. \ref{newfil}, we discuss a possible new class of optical filters
having the symmetry of the two-dimensional Euclidean group, and their
possible applications in Sec. \ref{appl}.
In Sec. \ref{nonor}, we show that the formalism developed in this paper
can accommodate the cases where the polarization coordinates are
squeezed or sheared.  The formalism can be extended also the case
where the polarization plane is not perpendicular to the direction of
propagation.

Since the Lorentz group is relatively new in optical sciences, we
present the four-by-four and two-by-two representations of this group
in Appendices A and B respectively.  Since $E(2)$ group plays the central
role in this paper, we give an introduction to this group in Appendix C.

\section{Formulation of the Problem} \label{formalism}

In studying polarized light propagating
along the $z$ direction, the traditional approach is to consider the $x$
and $y$ components of the electric fields.  Their amplitude ratio and
the phase difference determine the state of polarization.  Thus, we can
change the polarization either by adjusting the amplitudes, by changing
the relative phases, or both.  For convenience, we call the optical
device which changes amplitudes an ``attenuator'' and the device which
changes the relative phase a ``phase shifter.''

Let us write the electric field vector as
\begin{eqnarray}\label{cosine}
E_{x} &=& A \cos{\left(kz - \omega t + \phi_{1}\right)} , \cr
E_{y} &=& B \cos{\left(kz - \omega t + \phi_{2}\right)} ,
\end{eqnarray}
where $A$ and $B$ are the amplitudes which are real and positive numbers,
and $\phi_{1}$ and $\phi_{2}$ are the phases of the $x$ and $y$
components respectively.  This form is useful not only in classical
optics but also applicable to coherent and squeezed states of
light~\cite{knp91,chirkin93}.

The traditional language for this two-component light is the Jones-matrix
formalism which is discussed in standard optics textbooks~\cite{predo93}.
In this formalism, the above two components are combined into one column
matrix with the exponential form for the sinusoidal function.
\begin{equation}\label{expo1}
\pmatrix{E_{x} \cr E_{y}} =
\pmatrix{A \exp{\left\{i(kz - \omega t + \phi_{1})\right\}}  \cr
B \exp{\left\{i(kz - \omega t + \phi_{2})\right\}}} .
\end{equation}
This column matrix is called the Jones vector.  The content of
polarization is determined by the ratio:
\begin{equation}
{E_{y}\over E_{x}} = \left({B\over A}\right) e^{i(\phi_{2} - \phi_{1})} .
\end{equation}
which can be written as one complex number:
\begin{equation}\label{ratio}
w = r e^{i\phi}
\end{equation}
with
$$
r = {B \over A} , \qquad \phi = \phi_{2} - \phi_{1} .
$$
The degree of polarization is measured by these two real numbers, which are
the amplitude ratio and the phase difference respectively.
The purpose of this paper is to discuss the transformation properties of
this complex number $w$.  The transformation takes place when the light
beam goes through an optical filter whose transmission properties are not
isotropic.

The text-book version of the Jones-matrix formalism~\cite{predo93}
starts with the projection operator:
\begin{equation}\label{projec}
\pmatrix{1 & 0 \cr 0 & 0} ,
\end{equation}
applicable to the Jones vector of Eq.(\ref{expo1}).  This operator
keeps the $x$ component and completely eliminates the $y$-component
of the electric field.  This is an oversimplification of the real
world where the attenuation factor in the $y$ direction is greater
than that along the $x$ direction.  We shall replace this projection
operator by an attenuation matrix which is closer to the real world.

Another element in the traditional formalism is
\begin{equation}\label{shif1}
P(0, \delta) = \pmatrix{e^{-i\delta/2} & 0 \cr 0 & e^{i\delta/2}} ,
\end{equation}
which leads to a phase difference of $\delta$ between the $x$ and $y$
components.  The polarization axes are not always the $x$ and $y$ axes.
For this reason, we need the rotation matrix
\begin{equation}\label{rot1}
R(\theta) = \pmatrix{\cos(\theta/2) & -\sin(\theta/2)
\cr \sin(\theta/2) & \cos(\theta/2)} .
\end{equation}
The traditional Jones-matrix formalism consists of systematic
combinations of the above three components given in Eq.(\ref{projec}),
Eq.(\ref{shif1}) and Eq.(\ref{rot1}).

In this paper, we replace the projection operator of Eq.(\ref{projec})
by a squeeze matrix.
There are two transverse directions which are perpendicular to each
other.  The absorption coefficient in one transverse direction could
be different from the coefficient along the other direction.  Thus,
there is the ``polarization'' coordinate in which the absorption can
be described by
\begin{equation}\label{atten}
\pmatrix{e^{-\eta_{1}} & 0 \cr 0 & e^{-\eta_{2}}} =
e^{-(\eta_{1} + \eta_{2})/2} \pmatrix{e^{\eta/2} & 0 \cr 0 & e^{-\eta/2}}
\end{equation}
with $\eta = \eta_{2} - \eta_{1}$ .  Let us look at the projection
operator of Eq.(\ref{projec}).  Physically, it means that the absorption
coefficient along the $y$ direction is much larger than along the $x$
direction.  The absorption matrix in Eq.(\ref{atten}) becomes the
projection matrix if $\eta_{1}$ is very close to zero and $\eta_{2}$
becomes infinitely large.  The projection operator of Eq.(\ref{projec})
is therefore a special case of the above attenuation matrix.

The attenuation matrix of Eq.(\ref{atten}) tells us that the electric
fields are attenuated at two different rates.  The exponential factor
$e^{-(\eta_{1} + \eta_{2})/2}$ reduces both components at the same rate
and does not affect the state of polarization.  The effect of
polarization is solely determined by the squeeze matrix
\begin{equation}\label{sq1}
S(0, \eta) = \pmatrix{e^{\eta/2} & 0 \cr 0 & e^{-\eta/2}} .
\end{equation}
This type of mathematical operation is quite familiar from studies of
squeezed states of light, if not from Lorentz boosts of spinors.  For
convenience, we call the above matrix an attenuator.  Thus, we are
expanding the Jones-matrix formalism by replacing the projection
operator of Eq.(\ref{projec}) by the squeeze operator in Eq.(\ref{sq1}).

The phase-shifter of Eq.(\ref{shif1}) can be written as
\begin{equation}
P(0,\delta) = \exp{\left(-i\delta J_{1} \right)} ,
\end{equation}
with
\begin{equation}\label{j1}
 J_{1} = {1 \over 2} \pmatrix{1 & 0 \cr 0 & -1} .
\end{equation}
Our notation for the Pauli sigma matrices is different from those
appearing in the conventional literature, and is explained in
detail in Appendix D.

The rotation operator of Eq.(\ref{rot1}) takes the form:
\begin{equation}
R(\theta) = \exp{\left(-i\theta J_{3} \right)} ,
\end{equation}
with $$   J_{3} = {1 \over 2} \pmatrix{0 & -i \cr i & 0} . $$
The squeeze operator of Eq.(\ref{sq1}) can also be written in the
exponential form:
\begin{equation}
S(0,\eta) = \exp{\left(-i\eta K_{1} \right)} ,
\end{equation}
with $$   K_{1} = {i \over 2} \pmatrix{1 & 0 \cr 0 & -1} . $$

It is now possible to construct a closed set of commutation relations
with the above generators $J_{1}$, $J_{3}$, and $K_{1}$.  Then the
result is a set of six generators given in Appendix B.  They are the
generators for the group $SL(2,c)$ which is locally isomorphic to the
six-parameter Lorentz group.  This is the mathematical content of this
paper.  Let us next exploit the physical contents of this mathematical
formalism.

\section{Combined Effects}\label{combine}

If the polarization coordinate is the same as the $xy$ coordinate where
the electric field components take the form of Eq.(\ref{cosine}), the
above attenuator is directly applicable to the column matrix of
Eq.(\ref{expo1}).  If the polarization coordinate is rotated by an angle
$\theta/2$, or by the matrix
\begin{equation}
R(\theta) = \pmatrix{\cos(\theta/2) & -\sin(\theta/2)
\cr \sin(\theta/2) & \cos(\theta/2)} ,
\end{equation}
then the squeeze matrix becomes
\widetext
\begin{eqnarray}\label{sq2}
S(\theta, \eta) &=& R(\theta) S(0, \eta) R(-\theta) \nonumber  \\[2mm]
&=& \pmatrix{e^{\eta/2}\cos^{2}(\theta/2) + e^{-\eta/2}\sin^{2}(\theta/2)
& (e^{\eta/2} - e^{-\eta/2})\cos(\theta/2) \sin(\theta/2)
\cr (e^{\eta/2} - e^{-\eta/2})\cos(\theta/2) \sin(\theta/2)
& e^{-\eta/2}\cos^{2}(\theta/2) + e^{\eta/2}\sin^{2}(\theta/2)} .
\end{eqnarray}

\narrowtext
We can obtain the inverse of this transformation by rotating the filter
or the above expression around the $z$-axis by 90 degrees.

If we apply two squeeze matrices, the net result becomes
\begin{equation}
S\left(\theta_{2}, \eta_{2}\right) S\left(\theta_{1}, \eta_{1}\right)
= S\left(\theta_{3}, \eta_{3}\right) R(\psi) ,
\end{equation}
where $R(\psi)$ is a rotation around the $z$ axis by $\psi$.  This means
that the multiplication of two squeeze matrices does not lead to another
squeeze matrix, but a squeeze matrix preceded by a rotation
matrix.  This aspect of the squeeze operation is well known from the
squeezed state of light, and has been discussed extensively in the
literature~\cite{knp91,chiao88,kitano89}.  As was noted in
Sec. \ref{formalism}, the generators $K_{1}$ and $J_{3}$ are needed for
the squeeze matrix of Eq.(\ref{sq2}).  The repeated application leads to
the commutation relation for these two generators:
\begin{equation}
\left[J_{3}, K_{1}\right] = i K_{2} .
\end{equation}
Indeed, $J_{3}, K_{1}$ and $K_{2}$ form a closed set of commutation
relations for the $Sp(2)$ or $O(2,1)$-like subgroup of
$SL(2,c)$~\cite{kitano89}.  This three-parameter subgroup has been
extensively discussed in connection with squeezed states of
light~\cite{yuen76,knp91}.

Another basic element is the optical filter with two different values
of the index of refraction along the two orthogonal directions.  The
effect on this filter can be written as
\begin{equation}\label{phase}
\pmatrix{e^{i\delta_{1}} & 0 \cr 0 & e^{i\delta_{2}}}
= e^{-i(\delta_{1} + \delta_{2})/2}
\pmatrix{e^{-i\delta/2} & 0 \cr 0 & e^{i\delta/2}} ,
\end{equation}
with $\delta = \delta_{2} - \delta_{1}$ .
In measurement processes, the overall phase factor
$e^{-i(\delta_{1} + \delta_{2})/2}$
cannot be detected, and can therefore be deleted.  The polarization
effect of the filter is solely determined by the matrix
\begin{equation}
P(0, \delta) = \pmatrix{e^{-i\delta/2} & 0 \cr 0 & e^{i\delta/2}} .
\end{equation}
This form was noted as one of the basic components of the Jones-matrix
formalism in Sec. \ref{formalism}.  This phase-shifter matrix appears
like a rotation matrix around the $z$ axis in the theory of rotation
groups, but it plays a different role in this paper.  We shall hereafter
call this matrix a phase shifter.

Here also, if the polarization coordinate makes an angle $\theta$ with
the $xy$ coordinate system, the phase shifter takes the form
\widetext
\begin{eqnarray}\label{shif2}
P(\theta, \delta) &=& R(\theta) P(0, \delta) R(-\theta) \nonumber \\[2mm]
& = & \pmatrix{e^{-i\delta/2}\cos^{2}(\theta/2) +
e^{i\delta/2}\sin^{2}(\theta/2) &
(e^{-i\delta/2} - e^{i\delta/2})\cos(\theta/2) \sin(\theta/2) \cr
(e^{-i\delta/2} - e^{i\delta/2})\cos(\theta/2) \sin(\theta/2)
& e^{i\delta/2}\cos^{2}(\theta/2) + e^{-i\delta/2}\sin^{2}(\theta/2)} .
\end{eqnarray}
\narrowtext
Here again, we can obtain the inverse of this transformation by rotating
the filter around the $z$-axis by 90 degrees.

If we consider only the phase shifters, the mathematics is
basically repeated applications of $J_{1}$ and $J_{2}$, resulting in
applications also of $J_{3}$, where their explicit two-by-two matrix
forms are given in Appendix B.  Thus, the phase-shift filters form an
$SU(2)$ or $O(3)$-like subgroup of the group $SL(2,c)$.

If we use both the attenuators and phase shifters, the result is the
full $SL(2,c)$ group with six parameters.  The transformation matrix is
usually written as
\begin{equation}\label{abgd1}
L = \pmatrix{\alpha & \beta \cr \gamma & \rho} ,
\end{equation}
with the condition that its determinant be one:
$\alpha\rho - \gamma\beta$ = 1.  The repeated application of
two matrices of this kind results in
\begin{eqnarray}\label{abgd2}
\pmatrix{\alpha_{2} & \beta_{2} \cr \gamma_{2} & \rho_{2}}
\pmatrix{\alpha_{1} & \beta_{1}
\cr \gamma_{1} & \rho_{1}} \hspace{30mm}  \nonumber \\[2mm]
\hspace{16mm} = \pmatrix{\alpha_{2}\alpha_{1} + \beta_{2}\gamma_{1} &
\alpha_{2}\beta_{1} + \beta_{2}\rho_{1} \cr
\gamma_{2}\alpha_{1} + \rho_{2}\gamma_{1} &
\gamma_{2}\beta_{1} + \rho_{2}\rho_{1}} .
\end{eqnarray}
The most general form of the polarization transformation is the
application of this algebra to the column matrix of Eq.(\ref{expo1}).

The generators of the rotation or phase-shifters are Hermitian.  Thus,
the unitary subset of the $L$ matrices of Eq.(\ref{abgd1}), it represents
the three-parameter rotation-like subgroup or the phase-shifters and
their repeated applications.  The generators $J_{3}$, $K_{1}$ and $K_{2}$
are all imaginary.  Thus, the real subset of the $L$ matrices represents
the attenuation filters and their repeated applications.

\section{Bilinear Conformal Representation of the Lorentz
Group}\label{conform}
In the present formulation of polarization optics, we are interested
in calculating one complex variable defined in Eq.(\ref{ratio}).
As was noted in Ref.~\cite{hkn96}, we can obtain the same algebraic
result by using the bilinear transformation :
\begin{equation}\label{bilin1}
w' = {\rho w + \gamma \over \beta w + \alpha} .
\end{equation}
The repeated applications of these two transformations can be achieved
from
\begin{equation}\label{bilin2}
w_{1} = {\rho_{1} w + \gamma_{1} \over \beta_{1} w + \alpha_{1}} , \quad
w_{2} = {\rho_{2} w_{1} + \gamma_{2} \over \beta_{2} w_{1} + \alpha_{2}} .
\end{equation}
Then, it is possible to write $w_{2}$ as a function of $w$, and the result
is
\begin{equation}\label{bilin3}
w_{2} = {(\gamma_{2}\beta_{1} + \rho_{2}\rho_{1} ) w +
(\gamma_{2}\alpha_{1} + \rho_{2}\gamma_{1})
\over (\alpha_{2}\beta_{1} + \beta_{2}\rho_{1}) w +
(\alpha_{2}\alpha_{1} + \beta_{2}\gamma_{1})} .
\end{equation}
This is a reproduction of the algebra given in the matrix multiplication
of Eq.(\ref{abgd2}).  The form given in Eq.(\ref{bilin1}) is the bilinear
representation of the Lorentz group~\cite{barg47}.

Let us consider the physical interpretation of this result.  If we
apply the matrix $L$ of Eq.(\ref{abgd1}) to the column vector of
Eq.(\ref{expo1}), then
\begin{equation}
\pmatrix{\alpha & \beta \cr \gamma & \rho} \pmatrix{E_{x} \cr E_{y}}
= \pmatrix{\alpha E_{x} + \beta E_{y} \cr \gamma E_{x} + \rho E_{y}} ,
\end{equation}
which gives
\begin{equation}
{E'_{y} \over E'_{x}} = {\gamma E_{x} + \rho E_{y} \over
\alpha E_{x} + \beta E_{y}} .
\end{equation}
In term of the physical quantity $w$ defined in Eq.(\ref{ratio}), this
formula becomes
\begin{equation}
w' = {\gamma + \rho w \over \alpha + \beta w} .
\end{equation}
This equation is identical to the bilinear form given in Eq.(\ref{bilin1}),
and the ratio $w$ can now be identified with the $w$ variable defined as
the parameter of the bilinear representation of the Lorentz group in the
same equation.  Indeed, the bilinear representation is clearly the natural
language for polarization optics.

Let us next consider subgroups.  For the phase-shifters, the subgroup is
represented by the unitary matrix
\begin{equation}
\pmatrix{\alpha & \beta \cr -\beta^{*} & \alpha^{*}} .
\end{equation}
This form is preserved in the bilinear representation.  As for the
attenuation subgroup, all the components are real, and this aspect is also
preserved in the bilinear representation.

Let us next consider another form of bilinear transformation.
\begin{equation}
w' = \frac{(\exp{(-i\delta/2)} w}{\exp{(i\delta/2)} + \beta w} .
\end{equation}
This is also a form-preserving transformation.  This form can be
transformed into the matrix form
\begin{equation}
\pmatrix{e^{-i\delta/2} & \beta \cr 0 & e^{i\delta/2}} .
\end{equation}
Let us study this sub-representation in detail in Sec. \ref{newfil}.

\section{New Filters}\label{newfil}
We should note at this point that the Lorentz group has another set of
three-parameter subgroups.  They are like the two-dimensional Euclidean
group.  Let us consider one of them, which is generated by the matrices
$J_{1}$, $N_{2}$ and $N_{3}$, with
\begin{equation}
N_{2} = J_{2} + K_{3} , \qquad   N_{3} = J_{3} - K_{2} ,
\end{equation}
where
\begin{equation}
N_{2} = \pmatrix{0 & 1 \cr 0 & 0},  \qquad
N_{3} = \pmatrix{0 & -i \cr 0 & 0} .
\end{equation}
These matrices satisfy the commutation relations:
\begin{equation}\label{e2com}
[J_{1}, N_{2}]  = i N_{3},  \quad [J_{1}, N_{3}] = -iN_{2} \quad
[N_{2}, N_{3}]  = 0 .
\end{equation}
They indeed form a closed set of commutation relations.  As shown in
Appendix C, these commutation relations are like those for the
two-dimensional Euclidean group consisting of two translations and one
rotation around the origin.  This
group has been studied extensively in connection with the space-time
symmetries of massless particles, where $J_{1}$ and the two $N$ generators
correspond to the helicity and gauge degrees of freedom
respectively~\cite{hks82}.

The physics of $J_{1}$ is well known through the phase shifter given in
Eq.(\ref{shif1}).  If the angle $\delta$ is $\pi/2$, the phase shifter
becomes a quarter-wave shifter, which we write as
\begin{equation}
Q = P(0, \pi/2) = \pmatrix{e^{-i\pi/4} & 0 \cr 0 & e^{i\pi/4}} .
\end{equation}
Then $J_{2}$ and $K_{3}$ are the quarter-wave conjugates of $J_{3}$ and
$K_{2}$ respectively:
\begin{equation}
J_{2} = Q J_{3} Q^{-1} , \qquad K_{3} = - QK_{2}Q^{-1} .
\end{equation}
Consequently,
\begin{equation}
N_{2} = Q N_{3} Q^{-1} .
\end{equation}
The $N$ generators lead to the following transformation matrices.
\begin{eqnarray}\label{tmatrix}
T_{2}(\tau) &=& \exp{(-i\tau N_{2})}
= \pmatrix{1 & i\tau \cr 0 & 1}, \nonumber \\[2mm]
T_{3}(\tau) &=& \exp{(-i\tau N_{3})} = \pmatrix{1 & -\tau \cr 0 & 1} .
\end{eqnarray}
It is clear that $T_{2}$ is the quarter-wave conjugate of $T_{3}$.
We can now concentrate on the transformation matrix $T_{3}$.

If $T_{3}$ is applied to the column matrix of Eq.(\ref{expo1}),
\begin{equation}\label{tau}
\pmatrix{1 & -\tau \cr 0 & 1} \pmatrix{E_{x} \cr E_{y}} =
\pmatrix{E_{x} - \tau E_{y} \cr E_{y}} .
\end{equation}
This new filter superposes the $y$ component of the electric field to the
$x$ component with an appropriate constant, but it leaves the $y$ component
invariant.

Let us examine how this is achieved.  The generator $N_{3}$ consists of
$J_{3}$ which generates rotations around the $z$ axis, and $K_{2}$ which
generates a squeeze along the $45^{o}$ axis.  Physically, $J_{3}$
generates optical activities.  Thus, the new filter consists of a
suitable combination of these two operations.  In both cases, we have
to take into account the overall attenuation factor.   This can be
measured by the attenuation of the $y$ component which is not affected
by the symmetry operation of Eq.(\ref{tau}).

Is it possible to produce optical filters of this kind?  Starting from
an optically active material, we can introduce an asymmetry in absorption
to it by either mechanical or electrical means.  Another approach would
be to pile up alternately the $J_{3}$-type and $K_{2}$-type layers.  In
either case, it is interesting to note that the combination of these two
effects produces a special effect predicted from the Lorentz group.

The $E(2)$-like symmetry includes transformations generated by $J_{1}$,
which is given in Eq.(\ref{j1}).  The transformation matrix is
\begin{equation}
P(0, \delta) = \pmatrix{e^{-i\delta/2} & 0 \cr 0 & e^{i\delta/2}} ,
\end{equation}
This matrix is given in Eq.(\ref{shif1}) and its physics are well
understood.  Let us apply this matrix to $T_{3}$ from left and from right.
Then
\begin{eqnarray}
P(0, \delta) T_{3}(\tau) &=&
\pmatrix{1 & - e^{-i\delta/2} \tau \cr 0 & 1} , \nonumber \\[3mm]
T_{3}(\tau) P(0, \delta) &=&
\pmatrix{1 & - e^{i\delta/2}\tau \cr 0 & 1} .
\end{eqnarray}
This leads to
\begin{eqnarray}
P(0, -\delta) T_{3}(\tau) P(0, \delta)
&=& T_{3}(e^{i\delta}\tau)  \nonumber \\[3mm]
&=& \pmatrix{1 & - e^{i\delta} \tau \cr 0 & 1} .
\end{eqnarray}
We can of course obtain the $P(0, -\delta)$ filter by rotating the
$P(0, \delta)$ around the $z$ axis by 90 degrees.
It is thus possible to add a phase factor to the $\tau$ variable using
phase shifters of the type $P(0, \delta)$.

\section{Possible Applications of the New Filter}\label{appl}
The three-dimensional rotation group occupies an important place in
many different branches of physics.  The group $O(2,1)$ also is useful
in a number of fields including optics~\cite{knp91,abra78,guil84,knp86}.
Thus, traditional attenuation and phase-shift filters may be useful in
constructing analog computers performing the symmetry operations of
these groups.

The group $E(2)$ is somewhat new in optics~\cite{ky92}.  However, as was
noted in Appendix C, it deals with translations and rotations on a flat
surface.  This filter may therefore be useful as a computational device
for recording and reading two-dimensional maps.  Furthermore, the $T_{3}$
matrix of Eq.(\ref{tmatrix}) has an interesting algebraic property:
\begin{equation}
\pmatrix{1 & \beta_{1} \cr 0 & 1}  \pmatrix{1 & \beta_{2} \cr 0 & 1}
= \pmatrix{1 & \beta_{1} + \beta_{2} \cr 0 & 1} .
\end{equation}
The matrix can therefore be used for converting multiplication into
addition, like the logarithmic function.  This is one of the most
basic operations in computational machines.

As for a more immediate application, let us consider lens optics.  It
is a trivial laboratory operation to rotate a given filter around the
$z$ axis by 90 degrees.  If we rotate the matrix of Eq.(\ref{beta}),
the result is the matrix of the form
\begin{equation}
\pmatrix{1 & 0 \cr \beta & 1} .
\end{equation}
This form together with the original form of Eq.(\ref{beta})
serve as lens and translation matrices respectively in para-axial
optics.  Indeed, a system of polarization filters can serve as an
analog computer for a multi-lens system.

Furthermore, the matrix of the form
\begin{equation}\label{beta}
\pmatrix{1 & \beta \cr 0 & 1}
\end{equation}
represents a ``shear'' transformation.  This is one of the
basic deformations in engineering applications.

The Lorentz group was introduced to physics as the basic language
for space-time symmetries of elementary particles~\cite{wig39}, but
it is becoming increasingly prominent in many branches of physics and
engineering including classical and quantum optics.  The optical
filters may provide excellent calculational tools for the Lorentz
group.  Thus, these filters may be useful as components of future
computers.

\section{Non-orthogonal Coordinate Systems}\label{nonor}
Since the light polarization is caused by anisotropic crystals,
the polarization coordinate is not always orthogonal.  Let us
consider first the case where the light polarization is along a
pair of skew or squeezed axes.

It was noted in this paper that the $O(2,1)$-like subgroup can take
care of attenuation filters.  A transformation matrix of this
subgroup can also transform the orthogonal coordinate system into a
squeezed coordinate system.  The matrix takes the form
\begin{equation}\label{skew}
\pmatrix{x' \cr y'} = \pmatrix{\cosh(\eta/2) & \sinh(\eta/2) \cr
\sinh(\eta/2) & \cosh(\eta/2)} \pmatrix{x \cr y} .
\end{equation}
The idea is to transform the squeezed coordinate system into the
orthogonal system using the matrix or its inverse given in the
above expression.  Next, we can perform the polarization algebra
developed in this paper for the orthogonal coordinate system.  We
then can transform the result obtained in the orthogonal system
back to the original squeezed coordinate system.

It is interesting to note that the transformation matrix given in
Eq.(\ref{skew}) is one of the transformation matrices within the
frame work of the Lorentz-group representation developed in this
paper, and there is no need to make the existing mathematics more
complicated.  The story is the same for the shear transformation
which takes the form
\begin{equation}
\pmatrix{x' \cr y'} = \pmatrix{1 & b \cr 0 & 1} \pmatrix{x \cr y} .
\end{equation}
This transformation is also well within the framework discussed in
this paper.

The polarization plane is not always perpendicular to the direction of
the propagation.  We can take care of this problem by extending our
two-by-two formalism of Eq.(\ref{abgd1}) into the three-by-three form
\begin{equation}
\pmatrix{\alpha & \beta & 0 \cr \gamma & \rho & 0 \cr 0 & 0 & 1}
\end{equation}
applicable to the $(x, y, z)$ coordinate system.  The polarization
plane can be rotated around the $x$ axis by
\begin{equation}
\pmatrix{1 & 0 & 0 \cr 0 & \cos\xi & -\sin\xi \cr 0 & \sin\xi &
\cos\xi} .
\end{equation}
Using this matrix or its inverse, we can bring the problem to the
orthogonal coordinate system.  After doing the standard polarization
algebra, we can go back to the original coordinate system.

\section*{Concluding Remarks}
In this paper, we have replaced the projection operator of
Eq.(\ref{projec}) by the squeeze matrix given in Eq.(\ref{sq1}).
This brings the Jones-matrix formalism closer to the real world.  This
also enables us to formulate the problem within the framework of the
Lorentz group.

There are now many powerful mathematical tools derivable from the
Lorentz group~\cite{knp91,knp86}.  We can use them to study more
systematically varous aspects of optics.  The present paper is the
first step toward a systematic exposition of polarization optics.
There are several interesting future problems.  First, it is now
possible to study the relation between the Jones vectors and the
Stokes parameters using the mathematical theorems of the Lorentz
group connecting Minkowskian four-vectors and $SL(2,c)$ spinors.
Furthermore, the natural language of the Lorentz group is based on
circularly polarized states of photons.  It is therefore an
interesting future problem to work out the Jones-matrix formalism
applicable to circularly polarized lights.

It is widely understood that the Lorentz group started gaining its
prominence in optics through squeezed states of light~\cite{yuen76}.
This is not true.  This group was discussed much earlier in connection
with polarization optics~\cite{barakat63}.  However, we learned the
connection between squeeze transformations and Lorentz boosts while
studying squeezed states of light~\cite{knp91}.  The Lorentz group
is now a powerful mathematical device because there are many squeeze
transformations in physics, including the Jones-matrix formalism
discussed in the present paper.

It is also gratifying to note that the symmetry of the Lorentz group
is useful for various engineering applications of light waves
including their polarizations, reflections, and their propagation
media~\cite{baum95}.  In physics, it is relatively new to study
Maxwell's equations in terms of the Lorentz group~\cite{knp86}.  It
is thus a challenging future problem to combine the symmetries of
electromagnetic waves developed in engineering with those developed
in physics.

\appendix

\section{Lorentz Transformations}
Let us consider the space-time coordinates $(x, y, z, t)$.  Then the
rotation around the $z$ axis is performed by the four-by-four matrix
\begin{equation}
\pmatrix{\cos\theta & -\sin\theta & 0 & 0 \cr \sin\theta & \cos\theta &
0 & 0 \cr 0 & 0 & 1 & 0 \cr 0 & 0 & 0 & 1} .
\end{equation}
This transformation is generated by
\begin{equation}
J_{3} = \pmatrix{0 & -i & 0 & 0 \cr i & 0 & 0 & 0 \cr
0 & 0 & 0 & 0 \cr 0 & 0 & 0 & 0} .
\end{equation}
Likewise we can write down the generators of rotations $J_{1}$ and
$J_{2}$ around the $x$ and $y$ axes respectively.  These three generators
satisfy the closed set of commutations relations
\begin{equation}\label{rota}
\left[J_{i}, J_{j} \right] = i \epsilon_{ijk} J_{k} .
\end{equation}
This set of commutation relations is for the three-dimensional rotation
group.

The Lorentz boost along the $z$ axis takes the form
\begin{equation}
\pmatrix{\cosh\eta & 0 & 0 & \sinh\eta \cr 0 & 1 & 0 & 0 \cr
0 & 0 & 1 & 0 \cr \sinh\eta & 0 & 0 & \cosh\eta} ,
\end{equation}
which is generated by
\begin{equation}
K_{3} = \pmatrix{0 & 0 & 0 & i \cr 0 & 0 & 0 & 0 \cr
0 & 0 & 0 & 0 \cr i & 0 & 0 & 0} .
\end{equation}
Likewise, we can write generators of boosts $K_{1}$ and $K_{2}$ along the
$x$ and $y$ axes respectively, and they take the form
\begin{equation}
K_{1} = \pmatrix{0 & i & 0 & 0 \cr i & 0 & 0 & 0 \cr
0 & 0 & 0 & 0 \cr 0 & 0 & 0 & 0} , \quad
K_{2} = \pmatrix{0 & 0 & i & 0 \cr 0 & 0 & 0 & 0 \cr
i & 0 & 0 & 0 \cr 0 & 0 & 0 & 0} .
\end{equation}
These boost generators satisfy the commutation relations
\begin{equation}\label{boosta}
\left[J_{i}, K_{j} \right] = i \epsilon_{ijk} K_{k} , \qquad
\left[K_{i}, K_{j} \right] = -i \epsilon_{ijk} J_{k} .
\end{equation}

Indeed, the three rotation generators and the three boost generators
satisfy the closed set of commutation relations given in Eq.(\ref{rota})
and Eq.(\ref{boosta}).  These three commutation relations form the
starting point of the Lorentz group.  The generators given in this
Appendix are four-by-four matrices, but they are not the only set
satisfying the commutation relations.   We can construct also six
two-by-two matrices satisfying the same set of commutation relations.
The group of transformations constructed from these two-by-matrices
is often called $SL(2,c)$ or the two-dimensional representation of
the Lorentz group.
Throughout the present paper, we used the two-by-two transformation
matrices constructed from the generators of the $SL(2,c)$ group.

\section{Spinors and Four-vectors in the Lorentz Group}
In Appendix A, we have noted that there are four-by-four and two-by-two
representations of the Lorentz group.  The four-by-four representation
is applicable to covariant four-vectors, while the two-by-two
transformation matrices are applicable to two-component spinors which
in the present case are Jones vectors.  The question then is whether
we can construct the four-vector from the spinors.  In the language of
polarization optics, the question is whether it is possible to
construct the coherency matrix~\cite{born80,perina71} from the Jones
vector.

With this point in mind, let us start from the following form of the
Pauli spin matrices.
\begin{eqnarray}
\sigma_{1} &=& \pmatrix{1 & 0 \cr 0 & -1} , \quad
\sigma_{2} = \pmatrix{0 & 1 \cr 1 & 0} , \nonumber \\[2mm]
\sigma_{3} &=& \pmatrix{0 & -i \cr i & 0} .
\end{eqnarray}
These matrices are written in a different convention.  Here  $\sigma_{3}$
is imaginary, while $\sigma_{2}$ is imaginary in the traditional notation.
Also in this convention, we can construct three rotation generators
\begin{equation}
J_{i} = {1 \over 2} \sigma_{i} ,
\end{equation}
which satisfy the closed set of commutation relations
\begin{equation}\label{comm1}
\left[J_{i}, J_{j}\right] = i \epsilon_{ijk} J_{k} .
\end{equation}
We can also construct three boost generators
\begin{equation}
K_{i} = {i \over 2} \sigma_{i} ,
\end{equation}
which satisfy the commutation relations
\begin{equation}\label{comm2}
\left[K_{i}, K_{j}\right] = -i \epsilon_{ijk} J_{k} .
\end{equation}
The $K_{i}$ matrices alone do not form a closed set of commutation
relations, and the rotation generators $J_{i}$ are needed to form a
closed set:
\begin{equation}\label{comm3}
\left[J_{i}, K_{j}\right] = i \epsilon_{ijk} K_{k} .
\end{equation}

The six matrices $J_{i}$ and $K_{i}$ form a closed set of commutation
relations, and they are like the generators of the Lorentz group applicable
to the (3 + 1)-dimensional Minkowski space.  The group generated by the
above six matrices is called $SL(2,c)$ consisting of all two-by-two complex
matrices with unit determinant.

In order to construct four-vectors, we need two different spinor
representations of the Lorentz group.  Let us go to the commutation
relations for the generators given in Eqs.(\ref{comm1}), (\ref{comm2}) and
(\ref{comm3}).  These commutators are
not invariant under the sign change of the rotation generators $J_{i}$,
but are invariant under the sign change of the squeeze operators $K_{i}$.
Thus, to each spinor representation, there is another representation with
the squeeze generators with opposite sign.  This allows us to construct
another representation with the generators:
\begin{equation}
\dot{J}_{i} = {1 \over 2} \sigma_{i}, \qquad
\dot{K}_{i} = {-i \over 2} \sigma_{i} .
\end{equation}
We call this representation the ``dotted'' representation.  If we write
the transformation matrix $L$ of Eq.(\ref{abgd1}) in terms of the
generators as
\begin{equation}
L = \exp\left\{-{i\over 2} \sum_{i=1}^{3}\left(\theta_{i}\sigma_{i} +
i\eta_{i}\sigma_{i}\right) \right\} ,
\end{equation}
then the transformation matrix in the dotted representation becomes
\begin{equation}\label{eldot}
\dot{L} = \exp\left\{-{i\over 2} \sum_{i=1}^{3}\left(\theta_{i}\sigma_{i}
- i\eta_{i}\sigma_{i}\right)\right\} .
\end{equation}
In both of the above matrices, Hermitian conjugation changes the
direction of rotation.  However, it does not change the direction of
boosts.  We can achieve this only by interchanging $L$ to $\dot{L}$,
and we shall call this the ``dot'' conjugation.

Likewise, there are two different set of spinors.  Let us use $u$ and
$v$ for the up and down spinors for ``undotted'' representation.  Then
$\dot{u}$ and $\dot{v}$ are for the dotted representation.  The
four-vectors are then constructed as~\cite{hks86}
\begin{eqnarray}
u\dot{u} &=& - (x - iy), \quad v\dot{v} = (x + iy), \cr
u\dot{v} &=& (t + z), \quad v\dot{u} = -(t - z)
\end{eqnarray}
leading to the matrix
\begin{equation}\label{dotmat}
C = \pmatrix{u \dot{v} & -u\dot{u} \cr v\dot{v} & -v\dot{u}}
   = \pmatrix{u \cr v} \pmatrix {\dot{v} & -\dot{u}} ,
\end{equation}
where $u$ and $\dot{u}$ are one if the spin is up, and are zero if the
spin is down, while $v$ and $\dot{v}$ are zero and one for the spin-up
and spin-down cases.
The transformation matrix applicable to the column vector in the above
expression is the two-by-two matrix given in Eq.(\ref{abgd1}).  What
is then the transformation matrix applicable to the row vector
$(\dot{v},~-\dot{u})$ from the right-hand side?  It is the transpose
of the matrix applicable to the column vector $(\dot{v},~-\dot{u})$.
We can obtain this column vector from
\begin{equation}\label{dotcol}
 \pmatrix {\dot{v} \cr -\dot{u}} ,
\end{equation}
by applying to it the matrix
\begin{equation}
g = -i\sigma_{3} = \pmatrix{0 & -1 \cr 1 & 0} .
\end{equation}
This matrix also has the property
\begin{equation}
g \sigma_{i} g^{-1} = -\left(\sigma_{i}\right)^{T} ,
\end{equation}
where the superscript $T$ means the transpose of the matrix.  The
transformation matrix applicable to the column vector of Eq.(\ref{dotcol})
is $\dot{L}$ of Eq.(\ref{eldot}).  Thus the matrix applicable to the row
vector $(\dot{v},~-\dot{u})$ in Eq.(\ref{dotmat}) is
\begin{equation}
\left\{g^{-1} \dot{L} g\right\}^{T} = g^{-1} \dot{L}^{T} g .
\end{equation}
This is precisely the Hermitian conjugate of $L$.

In optics, this two-by-two matrix form appears as the coherency matrix,
and it takes the form
\begin{equation}\label{cohm1}
C = \pmatrix{<E^{*}_{x}E_{x}> & <E^{*}_{y} E_{x}> \cr
<E^{*}_{x} E_{y}> & <E^{*}_{y} E_{y}>} ,
\end{equation}
where $<E^{*}_{i}E_{j}>$ is the time average of $E^{*}_{i}E_{j}$.
This matrix is convenient when we deal with light waves whose two transverse
components are only partially coherent.  In terms of the complex parameter
$w$, the coherency matrix is proportional to
\begin{equation}\label{cohm2}
C = \pmatrix{1 & r e^{i\delta} \cr
r e^{-i\delta} & r^{2}} ,
\end{equation}
if the $x$ and $y$ components are perfectly coherent with the phase
difference of $\delta$.  If they are totally incoherent, the
off-diagonal elements vanish in the above matrix.

Let us now consider its transformation properties.  As was noted by
Opartrny and Perina~\cite{perina93}, the matrix of Eq.(\ref{cohm1}) is
like
\begin{equation}\label{matC}
C = \pmatrix{t + z & x - iy \cr x + iy & t - z} ,
\end{equation}
where the set of variables $(x, y, z, t)$ is transformed like a four-vector
under Lorentz transformations.  Furthermore, it is known that the Lorentz
transformation of this four-vector is achieved through the formula
\begin{equation}\label{Ldag}
C' = L C L^{\dagger} ,
\end{equation}
where the transformation matrix $L$ is that of Eq.(\ref{abgd1}).  The
construction of four-vectors from the two-component spinors is not
a trivial task~\cite{hks86,bask95}.  The two-by-two representation of
Eq.(\ref{matC}) requires one more step of complication.

\section{Two-dimensional Euclidean Group}
Let us consider a two-dimensional plane and use the $xy$ coordinate
system.  Then $L_{z}$ defined as
\begin{equation}
L_{z} = - i\left\{x {\partial \over \partial y} -
y {\partial \over \partial x} \right\}
\end{equation}
will generate rotations around the origin.  The translation generators
are
\begin{equation}
P_{x} = -i {\partial \over \partial x} , \qquad
P_{y} = -i {\partial \over \partial y} .
\end{equation}
These generators satisfy the commutation relations:
\begin{equation}
[L_{z}, P_{x}] = i P_{y},  \qquad [L_{z}, P_{y}] = -iP_{x} \qquad
[P_{x}, P_{y}] = 0 .
\end{equation}
These commutation relations are like those given in Eq.(\ref{e2com}).
They become identical if $L_{z}$, $P_{x}$ and $P_{y}$ are replaced by
$J_{1}$, $N_{2}$ and $N_{3}$ respectively.

This group is not discussed often in physics, but is intimately related
to our daily life.  When we drive on the streets, we make translations
and rotations, and thus make transformations of this $E(2)$ group.  In
addition, this group reproduces the internal internal space-time
symmetry of massless particles~\cite{wig39}.  This aspect of the $E(2)$
group has been extensively discussed in the
literature~\cite{knp86,hks86,wein64}.

\end{document}